\documentstyle[11pt,aaspp4]{article}

\def\lsim{\lower.5ex\hbox{$\; \buildrel < \over \sim \;$}}
\def\gsim{\lower.5ex\hbox{$\; \buildrel > \over \sim \;$}}

\begin{document}
\tighten

\title{Diffuse 0.5-1 keV X-Rays and Nuclear Gamma-Rays from Fast 
Particles in the Local Hot Bubble}

\author{Vincent Tatischeff and Reuven Ramaty}
\affil{Laboratory for High Energy Astrophysics\\ Goddard Space 
Flight Center, Greenbelt, MD 20771}

\begin{abstract}

We show that interactions of fast particles with the boundary shell 
of the local hot bubble could make an important contribution to the 
0.5-1 keV diffuse X-ray background observed with ROSAT. The bulk of 
these nonthermal X-rays are due to line emission from fast O ions of 
energies around 1 MeV/nucleon. This is the typical energy per 
particle in the ejecta of the supernova which is thought to have 
energized the bubble. We find that there is sufficient total energy 
in the ejecta to produce X-rays of the required intensity, subject 
to the details of the evolution of the fast particle population 
since the supernova explosion (about 3$\times$10$^5$ years ago based 
on the age of the Geminga pulsar). The unequivocal signature of 
lines from deexcitations in fast ions is their large width ($\delta 
E/E$$\simeq$0.1 for O lines), which will clearly distinguishes them 
from X-ray lines produced in a hot plasma.

If a small fraction of the total ejecta energy is converted into 
accelerated particle kinetic energy ($\gsim$30 MeV/nucleon), the 
gamma-ray line emission produced in the boundary shell of the local 
hot bubble could account for the recently reported COMPTEL 
observations of nuclear gamma-ray lines from a broad region towards 
the Galactic center.

\end{abstract}


\eject

\section{INTRODUCTION}

The origin of the diffuse X-ray background in the 0.5-1.0 keV range 
is not well understood. While at lower energies the background is 
most likely Galactic due to thermal emission from both the local hot 
bubble (LHB) and the Galactic halo (Snowden et al. 1998), at higher 
energies it is thought to be mostly extragalactic (e.g. Fabian \& 
Barcons 1992). In the 0.5-1.0 keV energy range the X-ray background 
probably contains contributions from both the halo and extragalactic 
sources, but as the Galactic plane is optically thick in this energy 
range, these sources will be absorbed at low latitudes. Thus, the 
fact that the 0.5-1 keV background shows no strong absorption at low 
Galactic latitudes indicates that there are significant 
contributions in this energy range from relatively local Galactic 
sources (McCammon \& Sanders 1990). However, the contributions of 
the $\sim$10$^6$K plasma in the LHB (Snowden et al. 1998) and of 
stellar sources (Schmitt and Snowden 1990) are insufficient to 
account for all of the observed 0.5-1 keV background. 

In a previous paper we investigated in detail the X-ray line and 
continuum production from fast nonthermal ion interactions 
(Tatischeff, Ramaty, \& Kozlovsky 1998; see also Ramaty, Kozlovsky, 
\& Tatischeff 1997). We have shown that strong line emission is 
expected in the 0.5-1 keV energy range, mainly due to atomic 
deexcitations in fast O following electron capture and excitation. 
We applied these calculations to the Orion region from which nuclear 
gamma-ray lines were observed with COMPTEL (Bloemen et al. 1997a). 
In the present paper we investigate the contribution of nonthermal 
ion interactions in the LHB to the 0.5-1 keV diffuse X-ray 
background and calculate the nuclear gamma-ray line emission that is 
expected to accompany the X-rays.

\section{THE 0.5-1 KEV X-RAY DIFFUSE BACKGROUND}

We used the ${3 \over 4}$ keV (0.47-1.2 keV) ROSAT all-sky survey 
data (Snowden et al. 1995).  We have excluded regions from which we 
do not expect the observed X-rays to be produced predominantly by 
accelerated ions. We thus removed the quadrant 
270$^{\circ}$$<$b$^{II}$$<$60$^{\circ}$ which contains the Sco-Cen 
bubble and a possible Galactic X-ray bulge (Park et al. 1997a; 
Snowden et al. 1997), as well as the Orion-Eradinus bubble, the 
Cygnus bubble, and the brightest point sources in the Galactic 
plane. The remaining longitude-averaged emission is shown in 
Figure~1a (vertical bars) as a function of Galactic latitude. We see 
that the  distribution is quite smooth, with a mean count rate of 
$\sim$120$\times$10$^{-6}$ counts s$^{-1}$ arcmin$^{-2}$ at high 
latitudes, slightly decreasing near the Galactic plane. 

There are four established sources which contribute to the ${3 \over 
4}$ keV diffuse X-ray background (Figure~1a). We obtained the EXRB 
(extragalactic X-ray background) curve by extrapolating the 
power-law dependence of the background above 1 keV (Gendreau et al. 
1995) to the ${3 \over 4}$ keV band, and by taking into account 
photoelectric absorption employing the cross sections of Morrison \& 
McCammon (1983), solar abundances, and the average latitude 
dependence of the Galactic atomic hydrogen from the Bell Laboratory 
survey (Stark et al. 1992; see also Snowden et al. 1990). The 
existence of extended soft X-ray emission from the Galactic halo was 
recently established by Snowden et al. (1998) from X-ray shadowing 
observations in the ROSAT ${1 \over 4}$ keV (0.12-0.284 keV) band. 
The halo emission corresponds to a plasma temperature of 
$\sim$10$^{6}$ K and produces $\sim$1140$\times$10$^{-6}$ counts 
s$^{-1}$ arcmin$^{-2}$ and $\sim$410$\times$10$^{-6}$ counts 
s$^{-1}$ arcmin$^{-2}$ in the ${1 \over 4}$ keV band, for the north 
and the south polar regions respectively. We calculate the 
contribution of this plasma to the ${3 \over 4}$ keV band using the 
Raymond \& Smith (1977) thermal emission model and the ROSAT 
response function (Snowden et al. 1994). Assuming an average of 
1000$\times$10$^{-6}$ counts s$^{-1}$ arcmin$^{-2}$ in the ${1 \over 
4}$ keV band for latitudes b$^{II}$$>$65$^{\circ}$, we obtain 
20$\times$10$^{-6}$ counts s$^{-1}$ arcmin$^{-2}$ in the ${3 \over 
4}$ keV band at high latitudes. We derived the Galactic halo curve 
in Figure~1a by using this value and taking into account 
photoelectric absorption as for the EXRB.

The LHB is a cavity around the solar system filled with a rarefied 
($\lsim$5$\times$10$^{-3}$ cm$^{-3}$) $\sim$10$^{6}$ K plasma which 
accounts for all of the ${1 \over 4}$ keV diffuse emission at low 
latitudes (exclusive of discrete emission features). Cox \& Anderson 
(1982) have successfully modeled the observed ${1 \over 4}$ keV 
X-ray intensity by the reheating of a pre-existing cavity by a 
supernova blast wave containing $\sim$5$\times$10$^{50}$ ergs. It 
was suggested that the Geminga pulsar is the remnant of this 
supernova that occurred $\sim$3$\times$10$^5$ years ago, based on 
the age of the pulsar (Gehrels and Chen 1993). The radius of the LHB 
inferred from the ROSAT ${1 \over 4}$ keV observations ranges from 
about 40 to 90 pc along the Galactic plane and from about 50 to 130 
pc at high latitudes (Snowden et al. 1998). We approximate this 
cavity by an ellipsoid of revolution centered on the Sun with radius 
of 50 pc along the minor axis in the Galactic plane and 75 pc along 
the major axis perpendicular to the plane. In the displacement model 
(Sanders et al. 1977), which was shown to be consistent with the 
available data, the X-ray intensity is proportional to the radius of 
the cavity. Snowden et al. (1998) have established a scale factor of 
0.155 pc (10$^{-6}$ counts s$^{-1}$ arcmin$^{-2}$)$^{-1}$ for the 
${1 \over 4}$ keV band. Using again the Raymond \& Smith (1977) 
model and the ROSAT response function, we derived for this 10$^{6}$ 
K plasma a scale factor of 27.7 pc (10$^{-6}$ counts s$^{-1}$ 
arcmin$^{-2}$)$^{-1}$ in the ${3 \over 4}$ keV band. For our assumed 
ellipsoidal shape of the LHB (100 pc minor axis and 150 pc major 
axis) we obtain the LHB curve in Figure~1a which, as can be seen, 
contributes less than 3\% to the observed ${3 \over 4}$ keV band 
intensity. As the estimated average H column density within the LHB 
is less than $\sim$5$\times$10$^{18}$ cm$^{-2}$ (Bloch et al. 1986; 
Juda et al. 1991), we can safely neglect absorption for both the ${1 
\over 4}$ keV and ${3 \over 4}$ keV bands. We estimated the 
unresolved stellar contribution (the dM stars curve in Figure~1a) by 
scaling the curve in McCammon \& Sanders (1990, figure 8) to the 
ROSAT data (vertical bars in Figure~1a).

We see in Figure~1a that the total emission resulting from these 
four sources accounts for only $\sim$70\% of the observed intensity 
at high latitudes and for less than 20\% in the Galactic plane, thus 
requiring a significant contribution from sources within our Galaxy. 
The filled squares in Figure~1b show the required residual emission 
obtained by subtracting the total emission of the four established 
source distributions from the observed intensity in Figure~1a. While 
at high latitudes this residual emission could be extragalactic 
(e.g. Cen et al. 1995), or originate from the halo, a Galactic 
component is clearly needed to explain the residual emission in the 
plane. We now investigate the possibility that the excess ${3 \over 
4}$ keV emission is entirely Galactic resulting from fast ion 
interactions. 

\section{X-RAY AND GAMMA-RAY PRODUCTION}

We assume that the X-ray production takes place in the boundary 
shell of the LHB. The observed X-ray flux per unit solid angle can 
then be written as 
\begin{eqnarray} {d\Phi_{\rm x} \over d\Omega} = 
{1 \over 4\pi D^2} {dQ_{\rm x} \over d\Omega}. 
\end{eqnarray} 
\noindent Here $dQ_{\rm x}/d\Omega$ is the emission produced in a 
volume of the shell bounded by the solid angle element $d\Omega$ (as 
viewed from the Earth), and $D^2=\int_{r_1}^{r_2}q_{\rm x}r^2dr / 
\int_{r_1}^{r_2}q_{\rm x}dr$ is a mean square distance determined by 
the distribution of the X-ray emissivity $q_{\rm x}$ and the shell 
thickness $(r_2-r_1)$. We assume that at ${3 \over 4}$ keV 
absorption within the shell can be neglected. By further assuming 
that $dQ_{\rm x}/d\Omega$ is independent of the direction of 
observation $\Omega$, we obtain the solid curve in Figure~1b, where 
the root mean square distance $D$ was allowed to trace out the above 
defined ellipsoid (radii of 50 and 75 pc in the plane and 
perpendicular to the plane, respectively). This curve is, for now, 
arbitrarily normalized to the data.

As the bulk of the nonthermal ${3 \over 4}$ keV emission is line 
emission from fast O ions of typical energies around 1 MeV/nucleon 
(Tatischeff et al. 1998), we first consider a scenario in which the 
fast particles are the ejecta of the supernova which is thought to 
have energized the LHB. As the most efficient X-ray line production 
will result from fast particles enriched in heavy ions (mostly O), 
we assume that the progenitor of the supernova in question was a 
massive star which has lost its H and He envelope. We thus derive 
the spectrum of the fast ions below 2 MeV/nucleon using the velocity 
distribution of the ejected mass (2.85 M$_\odot$) of a supernova 
resulting from a 60 M$_\odot$ progenitor (Woosley, Langer, \& Weaver 
1993, figure 9). Above 2 MeV/nucleon, we extrapolate the particle 
spectrum with an $E^{-4}$ energy dependence (Fields et al. 1996), 
which is consistent with the ejected masses at speeds in excess of 
2$\times$10$^4$ and 3$\times$10$^4$ km s$^{-1}$ given by Woosley et 
al. (1993). The result is shown by the solid curve in Figure~2, 
where we plot the differential energy content in the ejecta, i.e the 
energy content per unit kinetic energy per nucleon $E$. Since the 
production of the nonthermal ${3 \over 4}$ keV X-rays is 
independent of any cutoff in the ejecta energy spectrum above 2 
MeV/nucleon, we extend the spectrum to infinity. 

We calculated the production of X-rays by fast ions with energy 
spectrum given by the SN60 curve in Figure~2 and composition of the 
ejecta given by Woosley et al. (1993, table 4). We allowed these 
ions to impinge upon the boundary shell of the LHB, assuming a 
steady state, thick target interaction model (Tatischeff et al. 
1998) with a neutral ambient medium of solar composition. The 
assumption that the X-rays are not absorbed in the shell requires 
that the H column depth through the shell be 
$\lsim$3$\times$10$^{20}$ cm$^{-2}$. Such a H column is sufficient 
to slow down a 2 MeV/nucleon O to about 1 MeV/nucleon. Since the 
bulk of the X-ray line emission from fast O ions takes place between 
0.4 and 2 MeV/nucleon (Tatischeff et al. 1998), the assumption of a 
thick target for the fast particles is not unreasonable because 
magnetic fields within the shell could confine the particles to 
longer paths than simple straight line trajectories through the 
shell. We used the SN60 spectrum in Figure~2 with a renormalization 
to a power deposition $\dot W$ such that the calculated ${3 \over 
4}$ keV count rate for a distance $D$=50 pc [Eq.~(1)] in the 
Galactic plane equals 60$\times$10$^{-6}$ counts s$^{-1}$ 
arcmin$^{-2}$ (the required emission to account for this background 
in the plane, Figure~1b). We obtain $\dot W$=2.6$\times$10$^{38}$ 
erg s$^{-1}$, which is the power required to produce the total X-ray 
emission in the shell $\int (dQ_{\rm x}/d\Omega) d\Omega$ [see 
Eq.~(1)]. This power integrated over 2$\times$10$^5$ years 
(comparable to the age of the Geminga pulsar) yields a total energy 
equal to the initial energy of the ejecta, 
$\sim$1.5$\times$10$^{51}$ erg. We thus conclude that it is possible 
that the required fraction of the ${3 \over 4}$ keV background 
(Figure~1) is indeed due to nonthermal interactions of the supernova 
ejecta.

The corresponding X-ray spectrum, shown in Figure~3, is obtained by 
using our nonthermal X-ray production code (Tatischeff et al. 1998). 
We see that the ${3 \over 4}$ keV band (0.47-1.2 keV) is dominated 
by the broad He-like O lines at 0.57 and 0.67 keV, with some 
contribution from H-like O at 0.65 keV. These lines, as well as the 
various other broad lines seen in Figure~3, are due to atomic 
deexcitations in the fast ions following electron capture and 
excitation. The narrow lines (e.g. the Fe lines at 6.4 and 7.06 
keV) are due to K-shell vacancy production in the ambient atoms by 
the fast ions. The continuum is due to inverse bremsstrahlung and 
secondary electron bremsstrahlung.   

We also calculated the gamma-ray line production in the 3-7 MeV 
energy band (Ramaty, Kozlovsky, \& Lingenfelter 1996) due to the 
fastest particles of the supernova ejecta, again assuming a thick 
target interaction model. In the Galactic plane we obtain 
$\Phi_{3-7}$=2.4$\times$10$^{-6}$ photons cm$^{-2}$ s$^{-1}$ 
sr$^{-1}$ which is much smaller than the COMPTEL flux 
($\sim$1.5$\times$10$^{-4}$ photons cm$^{-2}$ s$^{-1}$ sr$^{-1}$) 
from a broad region in the direction of the Galactic center (Bloemen 
et al. 1997b). We thus see that the fast particles of the supernova 
ejecta, while potentially capable of producing very interesting 
fluxes of X-ray line emission, cannot play any significant role in 
nuclear gamma-ray line production. This result also shows that 
supernova ejecta without further acceleration cannot lead to 
sufficient gamma-ray line production to account for the COMPTEL 
observations of Orion (Cameron et al. 1995; Fields et al. 1996; 
Tatischeff et al. 1996). This is because for the steep spectra of 
the ejecta the accompanying X-ray production would greatly exceed 
the ROSAT upper limits (Tatischeff et al. 1998). 

Since it is possible that the fast particles of the supernova ejecta 
lose their energy on a time scale shorter than $\sim$3$\times$10$^5$ 
years (the elapsed time since the supernova explosion), we also 
consider an alternative scenario in which both the X-rays and gamma 
rays are produced by accelerated particles. For generality, we 
specify neither the acceleration mechanism nor the acceleration time 
or site. We simply assume the spectrum that was used to calculate 
gamma-ray line emission in Orion (Ramaty et al. 1996), 
\begin{eqnarray}
{d\Phi \over dE} \propto E^{-1.5}e^{-E/E_0}~, 
\end{eqnarray}
\noindent which could result from shock acceleration. Employing the 
same steady state, thick target interaction model (Ramaty et al. 
1996; Tatischeff et al. 1998) as we used for the supernova ejecta 
spectrum, we calculate both the X-ray and gamma-ray emissions using 
Eq.~(2) and the SN60 ejecta composition (Woosley et al. 1993). The 
solid curve in Figure~4 shows the 3-7 MeV nuclear gamma-ray line 
flux that is expected to accompany the X-ray production that yields 
60$\times$10$^{-6}$ counts s$^{-1}$ arcmin$^{-2}$ in the ROSAT PSPC 
${3\over 4}$ keV band. The dashed horizontal line is the COMPTEL 
flux from the direction of the Galactic center, mentioned above. 
Taking this flux as an upper limit, we conclude that if the ${3\over 
4}$ keV X-ray background were due to accelerated particles 
[Eq.~(2)], their energy spectrum must be soft enough ($E_0\lsim$10 
MeV/nucleon) in order not to overproduce the nuclear gamma-ray line 
emission. The dashed curve in Figure~4 is the power deposition $\dot 
W$ that accompanies X-ray production in the entire boundary shell 
leading to a ${3\over 4}$ keV count rate of 60$\times$10$^{-6}$ 
counts s$^{-1}$ arcmin$^{-2}$ for a distance of 50 pc. We see that 
the most efficient X-ray production is for 2$\lsim$$E_0$$\lsim$10 
MeV/nucleon for which $\dot W$$\simeq$2$\times$10$^{38}$ erg 
s$^{-1}$, essentially the same as the power deposition for the 
supernova ejecta spectrum.

Finally we consider a hybrid model incorporating both the fast 
particles of the supernova ejecta and accelerated particles with 
spectrum given by Eq.~(2). We have normalized the supernova ejecta 
spectrum so that it by itself yields the required ROSAT ${3 
\over 4}$ keV count rate for a distance of 50 pc. This implies the 
power deposition of 2.6$\times$10$^{38}$ erg s$^{-1}$ given above. 
We then normalized the shock acceleration spectrum to a power 
deposition equal to 20\% of this value. For the same integration 
time of 2$\times$10$^5$ years used above, we obtain 
the shock acceleration spectra shown by the dashed curves in 
Figure~2.

The sum of the contributions from the two spectral components are 
shown in Table~1, for $E_0$=10, 30 and 100 MeV/nucleon. The $E_0$=0 
case represents only the supernova ejecta contribution without 
additional particle acceleration and $E_0$=100 MeV/nucleon is an 
upper limit derived for Orion (Tatischeff, Ramaty, Mandzhavidze 
1997) using EGRET gamma-ray data. We see that, while the 
contribution of the shock spectrum to the X-ray line production is 
small, it dominates the gamma-ray line production. Thus, for 
$E_0$=100 MeV/nucleon we can account for the observed COMPTEL flux 
with a power deposition of only 5.2$\times$10$^{37}$ erg s$^{-1}$ in 
the entire boundary shell. But if the nuclear gamma-ray lines are 
produced in only part of the boundary shell (1.5 sr, Bloemen et al. 
1997b) the power deposition could be as small as 2\% of that 
required for the production of the ${3 \over 4}$ keV background. 

\section{DISCUSSION AND CONCLUSIONS}

We have shown that X-rays produced by fast ions could make an 
important contribution to the 0.5-1 keV diffuse X-ray background. 
The bulk of the emission is this energy range is due to line 
emission following electron capture and excitation in fast O ions of 
$\sim$1 MeV/nucleon. A promising source for these particles is the 
ejecta of the supernova which energized the LHB, and conceivably 
gave rise to the Geminga pulsar (Gehrels \& Chen 1993). The 
unequivocal signature of lines from deexcitations in fast ions is 
their large width ($\delta E/E$$\simeq$0.1 for O lines), which could 
clearly distinguish them from the X-ray lines produced in a hot 
plasma. O line emission has already been observed in the diffuse 
X-ray background (e.g. Rocchia et al. 1984), but the resolution of 
the solid-state detectors ($\sim$150 eV FWHM) is not sufficient to 
make the distinction.

For the calculation of the X-ray production we adopted a model in 
which the fast ions interact in the boundary shell of the LHB. Since 
this shell is probably more inhomogeneous than the hot plasma within 
the bubble, we expect significant spatial variation of the 0.5-1 keV 
background. Small-scale variations in the ${1 \over 4}$ keV 
background have already been observed, suggesting emission 
associated with the boundary shell (Park, Finley, \& Snowden 1997b). 
Further information on the validity of our model could be obtained 
from observations of X-ray shadows due to molecular clouds, such as 
the observations of Kuntz, Snowden, \& Verter (1997).

We have also calculated the gamma-ray line emission in the 3-7 MeV 
energy range, which is due mostly to nuclear deexcitations in C and 
O. We showed that the gamma-ray line emission produced by the fast 
supernova ejecta is unobservable with current instruments. However, 
if a significant fraction (at least 2\%) of the ejecta energy is 
converted into accelerated particle kinetic energy, the gamma-ray 
line emission produced in the local hot bubble could account for the 
recent COMPTEL observations (Bloemen et al. 1997b) from a broad 
region towards the Galactic center. Such local production could 
provide an explanation to the broad latitude extend of this 
emission. 

We wish to acknowledge useful discussions with Steve Snowden. V. T. 
acknowledges an NRC Research Associateship at the Goddard Space 
Flight Center.

\clearpage

\begin{deluxetable}{ccccccccc}
\tablecaption{Predicted X-ray and gamma-ray emissions at low 
latitudes\tablenotemark{a}}
\tablewidth{20pc}
\tablehead{
\colhead{$E_0$ (MeV/nucl)} & 
\colhead{X-ray\tablenotemark{b}} & 
\colhead{Gamma-ray\tablenotemark{c}} \\ }
\startdata
0\tablenotemark{d} & 60.0 & 0.024 \nl
10                 & 74.5 & 0.22  \nl
30                 & 71.0 & 0.70  \nl
100                & 67.3 & 1.43  \nl
\enddata
{\tablenotetext{a}{For a combination of the SN60 and shock spectra 
of Figure~2 and a distance to the emission source [Eq.~(1)] 
$D$=50 pc. The SN60 and shock spectra are normalized to powers 
of 2.6$\times$10$^{38}$ and 5.2$\times$10$^{37}$ erg s$^{-1}$, 
respectively.}
\tablenotetext{b}{In  units of 10$^{-6}$ counts s$^{-1}$ 
arcmin$^{-2}$ in the ROSAT PSPC ${3 \over 4}$ keV band.} 
\tablenotetext{c}{In units of 10$^{-4}$ photons cm$^{-2}$ s$^{-1}$ 
sr$^{-1}$ in the 3-7 MeV energy range.}
\tablenotetext{d}{Supernova ejecta without additional particle 
acceleration.}} 
\end{deluxetable}

\clearpage

\begin{figure}[t]   
\begin{center}     
\leavevmode 
\epsfxsize=12.cm 
\epsfbox{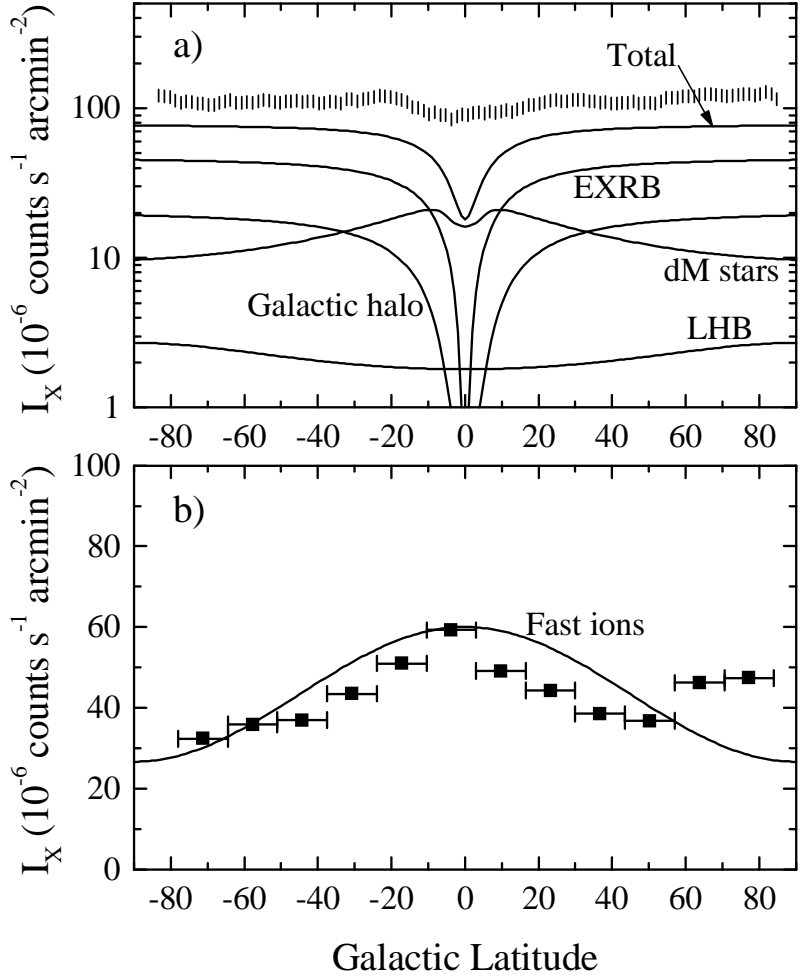}   
\end{center} 
\caption{Longitude averaged ${3 \over 4}$ keV diffuse X-ray 
background components as functions of Galactic latitude. Panel (a): 
vertical bars -- total ROSAT PSPC emission; total solid curve -- sum 
of contributions from the extragalactic X-ray background (EXRB), late-type 
dwarf stars, the Galactic halo and the local hot bubble. 
Panel (b): filled squares -- residual emission after subtraction 
of the total curve in panel (a) from the data, averaged over 
13.5$^\circ$ bins; solid curve -- X-ray emission from fast ion 
interactions with the boundary shell of the LHB (see text).} 
\label{fig:1} 
\end{figure}

\begin{figure}[t]   
\begin{center}     
\leavevmode 
\epsfxsize=17.cm 
\epsfbox{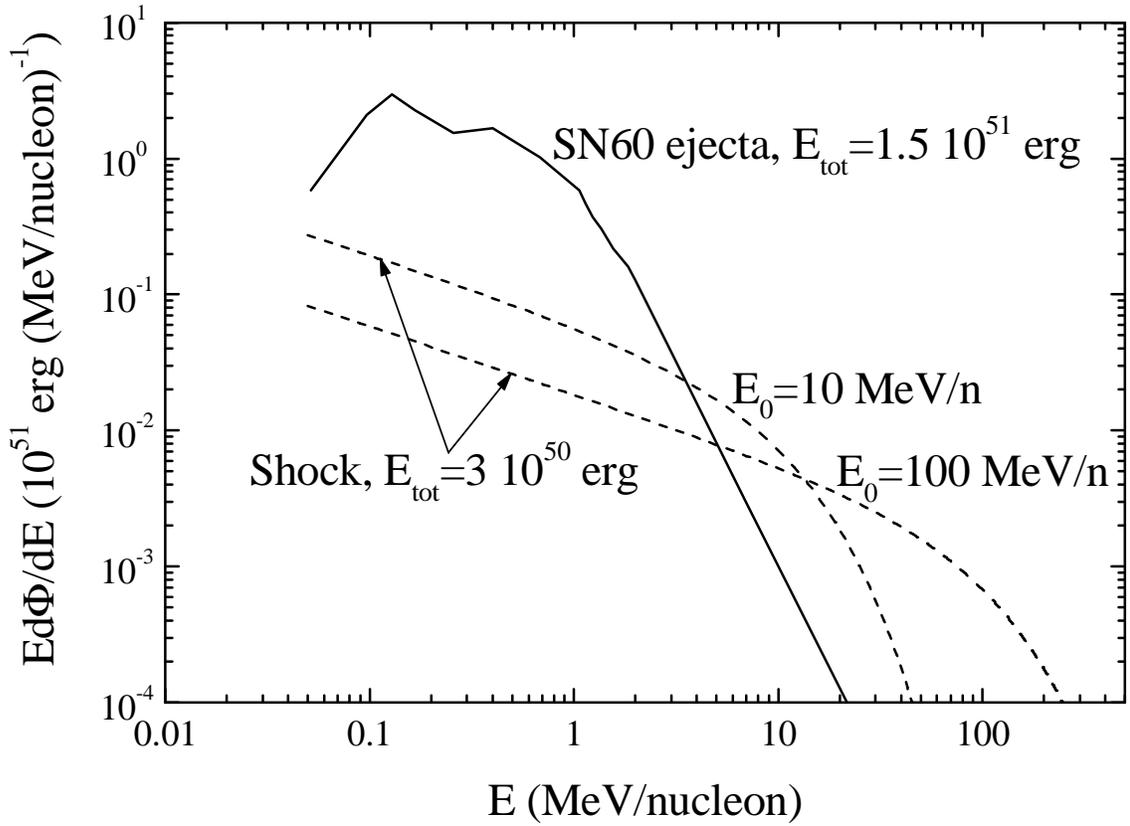}   
\end{center}   
\caption{Fast particle differential energy contents. Solid curve -- 
ejecta of a supernova resulting from a 60 M$_{\odot}$ progenitor 
(Woosley et al. 1993) containing 1.5$\times$10$^{51}$ erg; dashed 
curves -- shock acceleration spectra [Eq.~(2)] normalized to a total 
accelerated particle kinetic energy 
content of 20\% of that of the ejecta.}
\label{fig:2} 
\end{figure}

\begin{figure}[t]   
\begin{center}     
\leavevmode 
\epsfxsize=17.cm 
\epsfbox{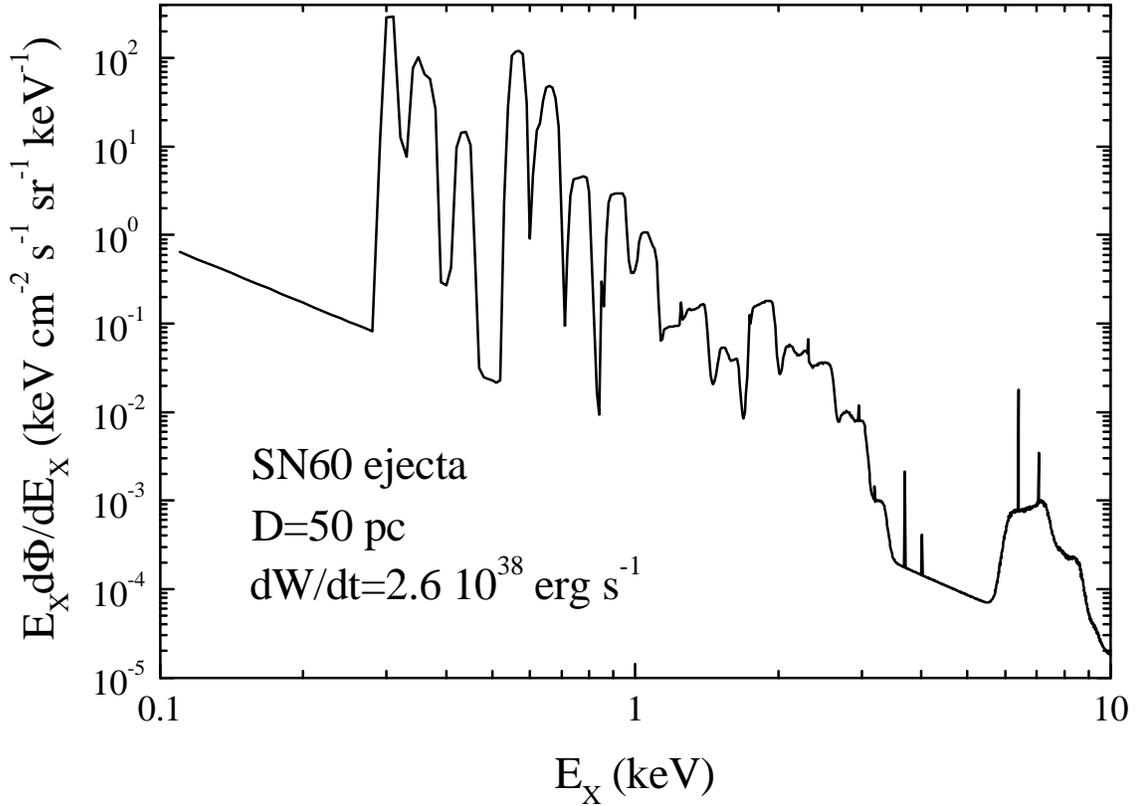}   
\end{center}   
\caption{X-ray emission produced by the ejecta of a supernova from a 60 M$_\odot$
progenitor (Woosley et al. 1993) interacting in the 
boundary shell of the LHB. The calculation is normalized to a total 
power of 2.6$\times$10$^{38}$ erg s$^{-1}$ deposited in the shell by 
the ejecta, yielding 60$\times$10$^{-6}$ counts s$^{-1}$ arcmin$^{-2}$ in 
the ROSAT PSPC ${3 \over 4}$ keV band for a distance of 50 pc.} 
\label{fig:3} 
\end{figure}

\begin{figure}[t]   
\begin{center}     
\leavevmode 
\epsfxsize=17.cm 
\epsfbox{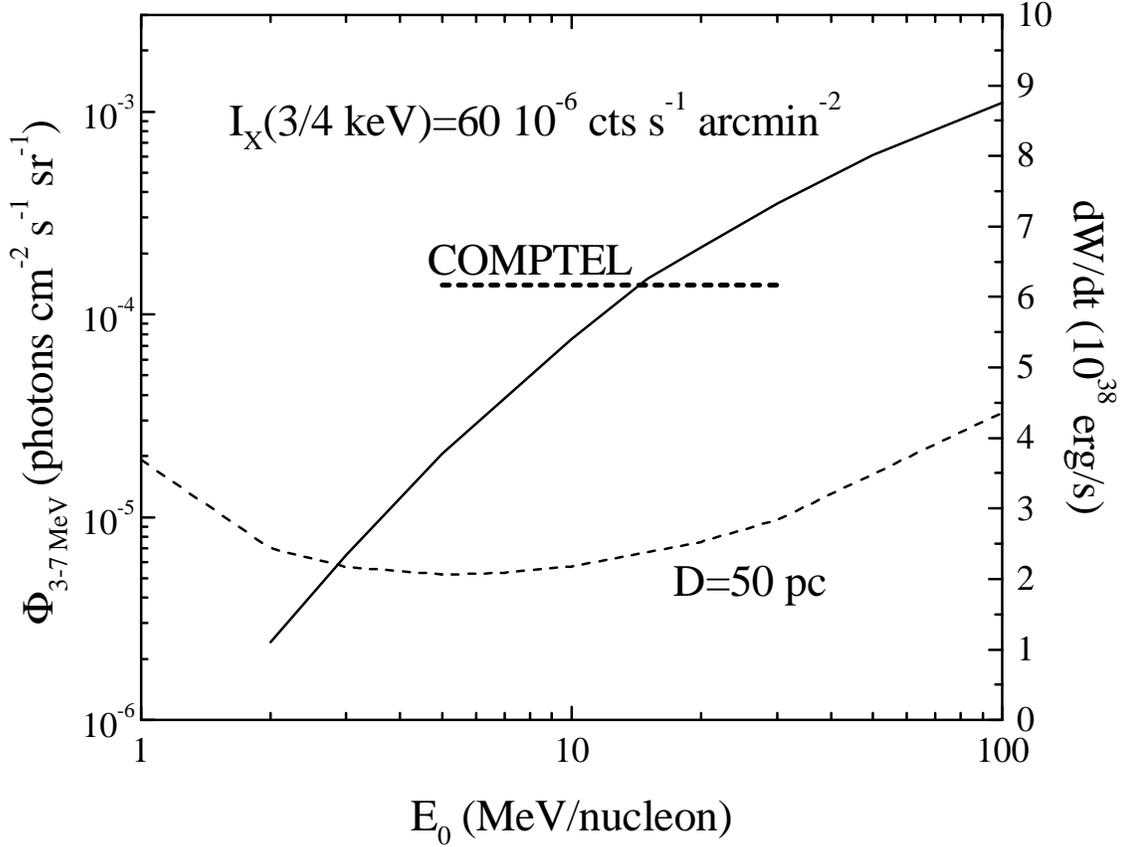}   
\end{center}   
\caption{Solid curve -- nuclear gamma-ray line emission in the 
3-7 MeV energy range (left vertical axis) that accompanies the 
X-ray production by the shock acceleration spectrum (Figure~2) 
as a function of $E_0$ [Eq.~(2)]. The calculation is normalized 
to an X-ray production of 60$\times$10$^{-6}$ counts s$^{-1}$ 
arcmin$^{-2}$ in the ROSAT PSPC ${3 \over 4}$ keV band. Also 
shown is the COMPTEL flux from a broad region in the direction 
of the Galactic center (Bloemen et al. 1997b). Dashed curve -- 
total power deposited in the shell by the accelerated ions 
(right vertical axis) as a function of $E_0$. The calculation 
is normalized to provide the same X-ray production in the ${3 
\over 4}$ keV band for a distance of 50 pc.} 
\label{fig:4} 
\end{figure}

\end{document}